\documentclass[letterpaper,aps,prb,twocolumn,showpacs,superscriptaddress]{revtex4}
\usepackage{graphicx}
\usepackage{amsmath}
\usepackage{epstopdf}
\usepackage{amsfonts}
\usepackage{amssymb}

\begin{document}

\title{Spin transport in mesoscopic rings with inhomogeneous spin-orbit coupling}
\author{Yaroslav Tserkovnyak}
\affiliation{Department of Physics and Astronomy, University of California, Los Angeles, California 90095, USA}
\affiliation{Centre for Advanced Study at the Norwegian Academy of Science and Letters, Drammensveien 78, NO-0271 Oslo, Norway}
\author{Arne Brataas}
\affiliation{Centre for Advanced Study at the Norwegian Academy of Science and Letters, Drammensveien 78, NO-0271 Oslo, Norway}
\affiliation{Department of Physics, Norwegian University of Science and Technology, NO-7491 Trondheim, Norway}

\begin{abstract}
We revisit the problem of electron transport through mesoscopic rings
with spin-orbit (SO) interaction. In the well-known path-integral approach, the scattering
states for a quasi-1D ring with quasi-1D leads can be expressed in terms of
spinless electrons subject to a fictitious magnetic flux. We show that
spin-dependent quantum-interference effects in small rings are strongest for spatially
inhomogeneous SO interactions, in which case spin currents can be controlled
by a small external magnetic field. Mesoscopic spin Hall effects in four-terminal
rings can also be understood in terms of the fictitious magnetic flux.
\end{abstract}

\date{\today}
\maketitle


\section{Introduction}

The Aharonov-Bohm (AB) effect in quasi-one-dimensional (1D) rings\cite{aharonovPR59} and, more generally, the adiabatic\cite{berryPRSLA84} and nonadiabatic\cite{aharonovPRL87} Berry holonomies manifest nontrivial quantum topology. Recently, this has attracted much attention in mesoscopic transport, exotic particle statistics, and topological quantum computation. The interest in spintronics, which aims to inject, manipulate, and detect electron spins in electronic devices, has also led many authors to revisit geometric aspects of transport in semiconductors with spin-orbit (SO) coupling,\cite{meirPRL89,mathurPRL92,qianPRL94} mainly focusing on variants of the Aharonov-Casher (AC) effect.\cite{aharonovPRL84} The AC effect in planar structures with Rashba SO coupling was recently observed in single rings \cite{konigPRL06} and ring arrays.\cite{kogaPRB06}

\begin{figure}[tbp]
\includegraphics[width=0.8\linewidth,clip=]{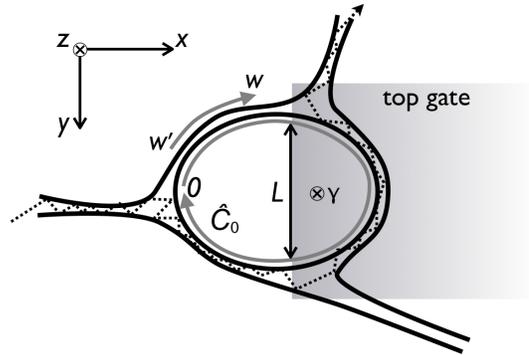}
\caption{A multiterminal quasi-1D ring in the $xy$ plane. A spatially inhomogeneous SO interaction is induced by the top gate. $\hat{C}_0$ defines SO-induced spin rotation in a full clockwise loop around the ring with respect to a chosen reference point $0$. The dotted line shows a particle trajectory between two leads with a winding number $n=-1$.}
\label{sc}
\end{figure}

Existing studies of AC related effects in mesoscopic rings focus on spatially homogeneous SO coupling.\cite{nittaAPL99,soumaPRL05,konigPRL06} However, we will show that quantum-interference effects in mesoscopic rings are stronger for spatially varying SO interactions in systems smaller than the SO precession length (typically 1~$\mu$m or longer in InAs-based heterostructures).\cite{brouwerPRB02,sternPRL92} An inhomogeneous SO interaction can be experimentally realized by electrostatic gates in semiconductor nanostructures, as sketched in Fig.~\ref{sc}. Electrostatic gates partially covering a planar ring induce an inhomogeneous macroscopic electric field perpendicular to the ring, influencing the electron motion via the spin-orbit interaction. Such an inhomogeneous SO coupling gives rise to a fictitious magnetic field with $SU(2)$ symmetry. In the weak SO limit, the corresponding fictitious flux dominates the geometric AC phase. Spin and charge flow in mesoscopic rings can therefore be manipulated by a combination of spatially varying SO interactions, which induces fictitious magnetic fields, and weak external magnetic fields.

Quantum-interference effects due to SO interactions linear in momentum in quasi-1D structures are most conveniently studied with path integrals, see e.g., Ref.~\onlinecite{chakravartyPRP86}. This formulation gives a clear physical
picture of geometric aspects of spin transport in mesoscopic
rings, allowing one to complement and generalize recent discussions on AC phases in multiterminal
rings.\cite{nittaAPL99,soumaPRL05,konigPRL06}
The path-integral approach recasts spin transport with
arbitrary scalar disorder and smooth SO coupling in terms of
conventional AB physics for spinless particles and purely geometric $SU(2)$
phase factors, and is valid, e.g., for both diffusive and ballistic
transport. Not being limited to two-terminal configurations, we will also see that the spin Hall effect in
four-terminal rings, such as that shown in Fig.~\ref{ex}, can be mapped onto the usual spinless Hall effect.

\section{Model}

We consider a ring in the $xy$ plane as shown in Fig.~\ref{sc}.
The electronic states are determined by the Hamiltonian 
\begin{equation}
\hat{H}=\frac{1}{2m}[-i\hbar \boldsymbol{\nabla }+(e/c)\mathbf{\hat{A}}(\mathbf{r})]^{2}-eV(\mathbf{r})\,,  \label{H}
\end{equation}
where $m$ is the effective mass, $-e$ electron charge, and the scalar
potential $V(\mathbf{r})$ includes applied, gate, and disorder fields. Hats
denote $2\times 2$ matrix structure in spin space for spin-$1/2$ particles.
In the following, we will assume a vacuumlike SO coupling (although the
arguments hold for any SO interaction linear in momentum, such as the
Dresselhaus SO coupling,\cite{dresselhausPR55}) which can be described by the effective vector
potential 
\begin{equation}
\mathbf{\hat{A}}(\mathbf{r})=\mathbf{A}_{m}(\mathbf{r})+\lambda (\mathbf{r})%
\mathbf{E}_{\mathrm{eff}}(\mathbf{r})\times \boldsymbol{\hat{\sigma}}\,,
\label{A}
\end{equation}%
where $\mathbf{A}_{m}$ is the vector potential related to the applied
magnetic field, $\mathbf{B}_{m}=\boldsymbol{\nabla }\times \mathbf{A}_{m}$
(disregarding Zeeman splitting), $\boldsymbol{\hat{\sigma}}$ is a vector of Pauli matrices, and $\lambda $ is a phenomenological
material-dependent SO parameter (in vacuum, $\lambda =-\hbar /4mc$, but it is many orders of magnitude larger in narrow-gap semiconductors such as GaAs and, especially, InAs).
$\mathbf{E}_{\mathrm{eff}}$ can consist of both the applied electric field and
the macroscopic band-structure contribution due to the crystal fields.
 Note that the $\lambda^2$ term in the Hamiltonian (\ref{H}) can be absorbed by a redefinition of $V(\mathbf{r})$, since $(\mathbf{E}\times\boldsymbol{\hat{\sigma}})^2=2E^2$.
Electromagnetic $U(1)$ gauge invariance for the magnetic component of the effective vector potential $\mathbf{\hat{A}}$ gives rise to the AB effect and the $SU(2)$ gauge symmetry for the
spin-dependent component causes the AC effect.\cite{frohlichRMP93}
It is useful to define an effective $2\times 2$ magnetic field,
\begin{equation}
\mathbf{\hat{B}}=\boldsymbol{\nabla }\times \mathbf{\hat{A}}=\mathbf{B}_{m}+\mathbf{\hat{B}}_{f}\,,
\end{equation}
which includes the fictitious SO contribution
$\mathbf{\hat{B}}_{f}$. Note that the fictitious $2\times 2$ magnetic field $\mathbf{\hat{B}}_{f}$ vanishes for a spatially homogeneous SO interaction.

For the Rashba SO coupling,\cite{bychkovJPC84} $\mathbf{E}_{\rm eff}$ is along the $z$ direction, and, after including the magnitude of $\mathbf{E}_{\rm eff}$ into the coefficient $\lambda$, the fictitious field and the associated flux through the ring are given by
\begin{align}
\label{BF}
\mathbf{\hat{B}}_{f}&=(\boldsymbol{\hat{\sigma}}\cdot \boldsymbol{\nabla}\lambda )\mathbf{z}\,,\\
\hat{\phi}_{f}&=\int_{A}d^{2}r(\mathbf{z}\cdot \mathbf{\hat{B}}_{f})\,.
\label{FF}
\end{align}
Diagonalization of this flux in spin space determines the spin 
quantization axis with
corresponding fictitious magnetic fluxes of equal magnitude but opposite
sign for spins up and down. For
example, if $\lambda $ is induced by a top gate at $x>x_{0}$, with $\lambda
(x>x_{0})=\lambda _{0}$ under the gate and vanishing outside of the gate,
then $\hat{\phi}_{f}=\lambda _{0}L\hat{\sigma}_{x}$, where $L$ is the length of
the top gate edge at $x=x_{0}$ overlapping the loop area, see Fig.~\ref{sc}.
We may now na{\"i}vely conclude that the spin-up (down) transport is
governed by the transport coefficients $g(\phi_{m}\pm \gamma\phi_0)$, with the
spin-quantization axis along the $x$ axis.  This will be indeed justified below. $g(\phi)$ can denote any transport coefficients in two- or multiterminal configurations, e.g., the current
response in one contact to voltages applied at two other
contacts, as a function of the magnetic flux $\phi $ through the
ring, in the absence of the SO coupling. $\phi_{m}$ is the physical
magnetic field contribution to the flux, while $\gamma =\lambda _{0}L/\phi_0$ is
the fictitious contribution due to the SO interaction, in units of the magnetic flux quantum $\phi_0=hc/e$. Keeping only linear
in SO coupling $\gamma $ terms, the total conductance per spin equals\cite{meirPRL89} 
\begin{equation}
g_{c}=\frac{1}{2}[g(\phi_{m}+\gamma\phi_0)+g(\phi_{m}-\gamma\phi_0)]\approx g(\phi_{m})\,,
\label{gc}
\end{equation}
while the spin conductance is 
\begin{equation}
g_{s}=\frac{1}{2}[g(\phi_{m}+\gamma\phi_0)-g(\phi_{m}-\gamma\phi_0)]\approx\gamma\phi_0\partial_{\phi}g(\phi_{m})\,.  \label{gs}
\end{equation}
According to Eqs.~(\ref{gc}) and (\ref{gs}), the magnetic
flux $\phi_{m}$ through the ring can be used to modulate both spin and
charge transport and the spin response is proportional to the SO
coupling. Even such a simple setup, an electrostatic gate partially covering the ring and inducing inhomogeneous SO coupling, can thus exhibit nontrivial spintronic applications.

Note that the fictitious magnetic field (\ref{BF}) and the corresponding flux (\ref{FF}) are relevant only when we are interested in the linear in SO coupling effects. In particular, a homogeneous SO coupling $\lambda$ induces no fictitious field (\ref{BF}). Nevertheless, the noncommutative $2\times 2$ SO vector potential $\mathbf{\hat{A}}(\mathbf{r})$ does not allow a removal of the SO field by a gauge transformation altogether. The vanishing field (\ref{BF}) for homogeneous SO interactions suggests that the SO effects are manifested in transport properties at higher orders in the SO strength, as detailed in the following. In particular, even in the absence of a fictitious field $\mathbf{\hat{B}}_f$, there is in general a finite effective flux $\gamma$, which in the leading order is quadratic in the SO strength. Consequently, for a weak SO strength, inhomogeneous SO interactions are more effective than their homogeneous counterpart in generating spin-dependent transport effects. The relevant small parameter here is the fictitious magnetic flux $\gamma$, in units of $\phi_0$. In the above example, $\gamma=L/2l_{\rm so}$, where
\begin{equation}
l_{\rm so}=\frac{\phi_0}{2\lambda}=\frac{\pi\hbar^2}{\alpha m}
\label{lso}
\end{equation}
is the spin-precession length expressed in terms of the Rashba SO parameter $\alpha=e\hbar\lambda/cm$. Taking $\alpha=10^{-11}$~eV~m (which is at the upper limit of typical values for InAs-based heterostructures \cite{kogaPRB06}) and $m=0.05m_e$, in terms of the free-electron mass $m_e$, we get $l_{\rm so}\approx1/2$~$\mu$m. We thus conclude that linear in SO coupling effects will dominate in submicron systems if the SO profile is strongly inhomogeneous on the scale of the spin-precession length.

\section{Path-integral approach}
\label{pia}

For a more systematic study of topological properties for arbitrary SO strength, we express the
evolution operator in terms of path integrals. To this
end, we are interested in the $2\times 2$ propagator $\hat{K}(\mathbf{r},%
\mathbf{r}^{\prime };t)$ for the Schr{\"{o}}dinger equation $i\hbar \partial
_{t}\hat{\varphi}=\hat{H}\hat{\varphi}$, where $\hat{\varphi}(\mathbf{r},t)$
is the spinor column:\cite{chakravartyPRP86}
\begin{equation}
\hat{K}(\mathbf{r},\mathbf{r}^{\prime };t)=T_{c}e^{-\frac{2\pi i}{\phi_0}\int_{c}d\mathbf{r}%
\lambda \left( \mathbf{E}_{\mathrm{eff}}\times \boldsymbol{\hat{\sigma}}%
\right)}\int \mathcal{D}[\mathbf{r}(t)]e^{\frac{i}{\hbar}S}\,.   \label{K}
\end{equation}
Here, $\mathcal{D}[\mathbf{r}(t)]$ schematically denotes all trajectories in
space-time connecting points $\mathbf{r}$ and $\mathbf{r}^{\prime }$ in time 
$t$, $S[\mathbf{r}(t)]$ is the corresponding classical action for spinless
motion, and $T_{c}$ is the contour ordering operator that moves operators in
the expanded exponential which are further along the contour to the left.
The contour ordering is necessary for inhomogeneous effective fields $%
\mathbf{E}_{\mathrm{eff}}$, since the Pauli matrices do not commute.
We now assume the potential $V(\mathbf{r})$ constrains the orbital motion
within a narrow quasi-1D ring connected to several wires, see Fig.~\ref{sc}.
Within each wire and the ring, electrons can scatter and undergo
arbitrary orbital motion governed by the potential $V(\mathbf{r})$, but we disregard the net
spin precession determined by $\mathbf{E}_{\mathrm{eff}}$ for the
transverse motion within the quasi-1D channels, and focus on the phase
accumulated during the propagation along the ring. This requires the
characteristic SO precession length to be much longer than the wire widths,
which is easily realized experimentally. The key observation, according to
Eq.~(\ref{K}), is that the SO-induced phase is purely geometric and independent
of how fast the electrons propagate.

In order to formally separate the geometric contribution to the propagator (%
\ref{K}), we first need to make a convention for labeling quasi-1D paths
connecting an arbitrary point $w^\prime$ in the system (either
in the ring or the connected wires) to another point $w$. In the
following, we suppress the transverse degrees of freedom along the
connectors and the loop. The paths along the ring are classified according
to the number of clockwise windings, $n$. We define the $n=0$ path to be the
shortest clockwise path between the points $w^\prime$ and $%
w$. A finite positive (negative) $n$ corresponds to $n$
additional clockwise (counterclockwise) windings around the loop. The
shortest clockwise path from $w^\prime$ to $w$ is shown
in Fig.~\ref{sc}. In addition, we choose an arbitrary point in the loop [%
e.g., the contact between a wire and the ring, as in Fig.~\ref{sc}],
denoted by $w=0$, to be the reference point. Let us denote
the contour-ordered spin-rotation operator entering Eq.~(\ref{K}) for the $n$th
path from $w^\prime$ to $w$ by $\exp [i\hat{C}_{n}(%
w,w^\prime)]$, where $\hat{C}_{n}(w,w%
^\prime)$ is a $2\times 2$ Hermitian matrix determined by $\mathbf{E}_{%
\mathrm{eff}}$ along the path. The eigenstates $\epsilon$ of the Hamiltonian (\ref%
{H}) can now be found from the eigenvalue problem 
\begin{equation}
i\hbar \int dw^\prime\sum_{n}e^{i\hat{C}_{n}(w,w%
^\prime)}\partial _{t}K_{n}(w,w^\prime;t)\hat{\varphi}(w^\prime)=\epsilon \hat{\varphi}(w)
\label{E}
\end{equation}
at $t=0$, where $K_{n}(w,w^\prime;t)$ is the propagator for
spinless electron motion. The spin-orbit interaction contributes only to the
path-dependent $SU(2)$ geometric prefactor. The eigenvalue problem (\ref{E})
can be diagonalized in spin space, after we make several definitions: Let $\hat{C}_{0}=\hat{C}%
_{0}(0,0^{+})$, where $0^{+}$ is a point slightly clockwise offset from $0$,
so that $\hat{C}_{0}$ corresponds to one full cycle with respect to $w=0$. $%
e^{i\hat{C}_{0}}$ is diagonalized by a unitary transformation:
\begin{equation}
\hat{U}^{\dagger }e^{i\hat{C}_{0}}\hat{U}=\mathrm{diag}\{e^{\mp2\pi i\gamma}\}\,,
\end{equation}
with a unique $\gamma$ in the range $0\leq \gamma <1$. We next introduce a
fictitious vector potential $\mathbf{A}_{\gamma }(\mathbf{r})=-\mathbf{A}%
_{-\gamma }(\mathbf{r})$ corresponding to the magnetic flux $\pm \gamma $
(in units of $\phi _{0}$), respectively, through the loop, see Fig.~\ref{sc},
but in an otherwise arbitrary gauge. Finally,
\begin{equation}
\theta (w,w^{\prime})=\frac{2\pi}{\gamma\phi _{0}}\int_{w^{\prime }}^{w}d\mathbf{q}\cdot \mathbf{A}_{\gamma }
\end{equation}
is a line integral along the $n=0$ path from $w^{\prime }$ to $w$
around the loop, so that $\theta (w,w+0^{+})=2\pi $ and $\theta (w,w-0^{+})=0$
for points $w$ inside the ring. We are now ready to make the transformation: 
\begin{equation}
\hat{\varphi}(w)=e^{i\hat{C}_{0}(w,0)}\hat{U}\mathrm{diag}\{e^{\pm i\gamma
\theta (w,0)}\}\hat{\varphi}^{\prime }(w)\,.  \label{T}
\end{equation}%
Note that this is a smooth transformation when the SO interaction and the
magnetic field are smooth. Straightforward manipulations show that substituting Eq.~(\ref{T}) into Eq.~(\ref{E}) diagonalizes the eigenvalue problem for two spin species: 
\begin{equation}
\mathrm{diag}\{H_{\pm \gamma }\}\hat{\varphi}^{\prime }=\epsilon \hat{\varphi%
}^{\prime }\,,  \label{D}
\end{equation}%
where $H_{\pm \gamma }$ is the Hamiltonian with the potential $V(\mathbf{r})$%
, vector potential $\mathbf{A}_{m}(\mathbf{r})$ due to the external magnetic
field, but with the SO coupling replaced by the additional spin-dependent
fictitious vector potential $\mathbf{A}_{\pm \gamma }(\mathbf{r})$. The
eigenvalues and eigenstates of our original Hamiltonian (\ref{H}) can thus
be expressed in terms of the solution of the simpler problem, Eq.~(\ref{D}),
describing spinless electron experiencing a magnetic flux, which was discussed
extensively in various contexts. The spin texture corresponding to the SO
coupling is then added by a purely geometric unitary transformation (\ref{T}%
). SO interactions linear in momentum thus do not add any complexity to
the electronic structures of general multiterminal quasi-1D rings.
We thus remark that many of the results discussed in Ref.~\onlinecite{nittaAPL99}
can be readily obtained after solving the problem of magnetotransport for
spinless electrons and then making the transformation (\ref{T}) (valid for
arbitrarily strong SO coupling) in order to get detailed information about
spin as well as charge transport.

We can now return to justify the use of the fictitious magnetic field (\ref{BF}) for inhomogeneous SO coupling in the discussion leading to Eqs.~(\ref{gc}) and (\ref{gs}). In the
weak SO limit, to linear order in $\lambda$, 
\begin{equation}
e^{i\hat{C}_{0}}\approx e^{-2\pi i\frac{\hat{\phi}_{f}}{\phi _{0}}}\,,
\label{C}
\end{equation}%
where $\hat{\phi}_{f}$ is given by Eq.~(\ref{FF}). Using the Hausdorff
formula,
\begin{equation}
e^{\hat{A}}e^{\hat{B}}=e^{\hat{A}+\hat{B}+\frac{1}{2}[\hat{A},\hat{B}]+...}\,,
\end{equation}
it is clear that corrections to the approximation (\ref{C}) are
quadratic in $\lambda $, corresponding to non-commuting spin rotations along
the loop contour, and can be disregarded for systems smaller than the spin-precession length $l_{\rm so}$. The spin transport problem thus reduces to the AB effect for two spin
species along the quantization axis determined by $\hat{\phi}_{f}$, with
opposite fictitious fluxes. Since this quantum-interference effect is linear
in the SO strength, this regime can become useful in practice when a weak SO
coupling is modulated by external electrostatic gates. We should note here
that in the spirit of the approximation, i.e., for the spin-transport
properties linear in the SO strength, the additional spin transformation determined by $\hat{C%
}_{0}(r,0)$ in Eq.~(\ref{T}), which rotates spins at
position $\mathbf{r}$ by an angle linear in the SO coupling, can be disregarded, as well as ambiguities in
defining spin currents and spin conductances in the leads with a finite SO
coupling $\lambda $.

\section{Four-terminal longitudinal and Hall spin currents}

After the general discussion of Sec.~\ref{pia}, let us now consider a specific example that illustrates how the theory can be applied in practice to i) enhance spin-dependent effects and ii) control the size and direction of the induced spin currents and accumulations. We will analyze spin-dependent transport through a four-terminal conducting loop with spin-orbit interactions, using the theory of AB effect for coherent multiterminal conductors.\cite{buttikerPRL86}

\begin{figure}
\begin{center}\includegraphics[width=0.8\linewidth,clip=]{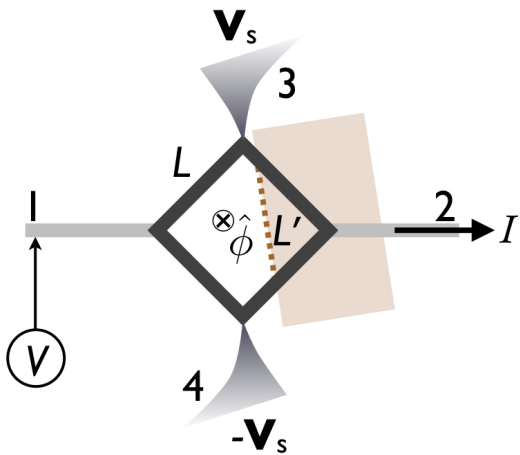}\end{center}
\caption{A diamond-shaped four-terminal loop. Voltage is applied to terminal 1, current is drawn out at 2 and Hall voltage and spin accumulation are measured at 3 and 4.}
\label{ex}
\end{figure}

Consider a diamond-shaped loop, contacted by four leads, as sketched in Fig.~\ref{ex}. Low-bias spinless transport between the leads, in the presence of a magnetic flux threading the loop, is fully determined by two functions $g(\phi)$ and $t(\phi)$ (that depends on microscopic details), supposing for simplicity the structure is mirror symmetric with respect to the axis connecting leads 1 and 3 as well as leads 2 and 4. $t(\phi)$ [$g(\phi)$] is the flux-dependent conductance relating the current in lead 2 (3) induced by voltage in lead 1, while three other leads are grounded. By symmetry, the conductance for lead 4 equals $t(-\phi)$, which is in general different from $t(\phi)$. The coefficient $g(\phi)=g(-\phi)$, on the other hand, is symmetric in magnetic field. For the ``transverse'' conductance $t(\phi)$, we define symmetric and antisymmetric components:
\begin{equation}
t_\pm(\phi)=\frac{1}{2}\left[t(\phi)\pm t(-\phi)\right]\,.
\label{sa}
\end{equation}

Let us apply a small voltage bias $V$ to lead 1, grounding lead 2, and calculate the current $I$ in lead 2 and voltages $V_{3}$ and $V_{4}$ induced in leads 3 and 4, respectively, assuming leads 3 and 4 are disconnected from any external circuitry so that $I_{3}=I_{4}=0$. We find
\begin{equation}
V_\pm(\phi)=\frac{t_\pm(\phi)}{2t_+(\phi)}V\,,
\end{equation}
where $V_{+}=(V_{4}+V_{3})/2$ and $V_{-}=(V_{4}-V_{3})/2$ is the Hall voltage. The induced current $I$ equals
\begin{equation}
I(\phi)=\left[g(\phi)+t_+(\phi)-\frac{t_-(\phi)^{2}}{t_+(\phi)}\right]V\,.
\label{IV}
\end{equation}
It therefore follows $I(\phi)=I(-\phi)$, as it should in effectively two-terminal conductor, according to time-reversal symmetry. The voltage difference $V_{-}(\phi)=-V_{-}(-\phi)$, on the other hand, is antisymmetric in magnetic field. This is an interference-induced Hall effect, in the absence of magnetic field within the wires.

Let us now return to the spin-orbit physics. Consider first a ring conductor with a uniform Rashba spin-orbit constant $\lambda$: $\hat{\mathbf{A}}=\lambda\mathbf{z}\times\hat{\boldsymbol{\sigma}}$. The effective magnetic flux is quadratic in $\lambda$ in this case, $\hat{\phi}_f\propto[\hat{A}_{x},\hat{A}_{y}]$:
\begin{equation}
\hat{\phi}_f\approx-\pi\left(\frac{L}{l_{\rm so}}\right)^{2}\phi_{0}\hat{\sigma}_{z}\,,
\label{phiz}
\end{equation}
where $l_{\rm so}$ is the spin-orbit precession length (\ref{lso}). We are assuming here and henceforth weak spin-orbit coupling on the scale set by the ring size: $L\ll l_{\rm so}$. The flux governing the AC effect on the two-terminal conductance is therefore $\phi_{s}=-\pi(L/l_{\rm so})^{2}\phi_{0}$. At the same time, a spin Hall effect develops, which is quadratic in the SO strength:
\begin{align}
V_{s4}&=V_{4}(\phi_{s})-V_{4}(-\phi_{s})=2V_{-}(\phi_{s})\nonumber\\
&=\frac{t_-(\phi_s)}{t_+(\phi_s)}V\approx\frac{\phi_s\partial_\phi t|_{\phi\to0}}{t}V\,,
\end{align}
where $V_{s4}=-V_{s3}$ are the $z$-axis spin accumulations induced in leads 3 and 4, respectively.

Next, suppose a top gate induces Rashba interaction only in a half-space. Let $L^{\prime}$ be the length of the gate edge overlapping the ring and $\mathbf{n}$ is the in-plane normal to the edge, pointing toward the region with a finite Rashba coupling $\lambda$, see Fig.~\ref{ex}. Using Eq.~(\ref{FF}), we get
\begin{equation}
\hat{\phi}_f/\phi_0=\frac{L^{\prime}}{2l_{\rm so}}(\hat{\boldsymbol{\sigma}}\cdot\mathbf{n})\,.
\end{equation}
Compare this result to Eq.~(\ref{phiz}). There are two important differences: Firstly, the spin-orbit inhomogeneity enhances the magnitude of the flux, making it linear in $\lambda$ rather than quadratic. Secondly, the spins are induced along the $\mathbf{n}$ direction determined by the edge orientation, rather than the 2DEG normal ($z$ axis) as in Eq.~(\ref{phiz}). Both AC and spin Hall effects discussed for the uniform $\lambda$ are similar in the present case, once we identify the new effective flux magnitude $\phi_{s}=(L^{\prime}/2l_{\rm so})\phi_{0}$ and the new spin quantization axis $\mathbf{n}$.

Finally, we note that the importance of the fictitious magnetic field (\ref{BF}) and the relation to the conventional Hall effect for the semiclassical boundary spin Hall physics was discussed in Ref.~\onlinecite{tserkovPRB07}, in a related but different context.

\section{Summary}

In conclusion, we have shown that a spatially inhomogeneous SO interaction enhances the spin-interference effects in rings smaller than the spin-precession length. Transport can be understood in terms of the AB physics with fictitious spin-dependent magnetic fluxes. Spin injection in two-terminal rings and spin Hall effect in four-terminal rings are enhanced and controlled by the edge of the SO interaction inhomogeneity.

\acknowledgments

This work was supported in part by the Research Council of Norway through
Grant Nos. 158518/143, 158547/431, and 167498/V30.

\end{document}